\documentclass[aps,prl,twocolumn,groupedaddress]{revtex4}
\usepackage{graphicx} 
\usepackage{amssymb} 
\begin{document}  

\title{Differential Entropy on Statistical Spaces}

\author{Jacques Calmet}  
\affiliation{ 
Institute for Algorithms and Cognitive Systems (IAKS)\\
University of Karlsruhe (TH), D-76131 Karlsruhe, Germany}
\email{calmet@ira.uka.de}

\author{Xavier Calmet}
\affiliation{Department of Physics and Astronomy \\  University of North Carolina at Chapel Hill, Chapel Hill,
NC 27599, USA}
\thanks{The work of X.C. was supported in part by the US Department of
Energy under Grant No. DE-FG02-97ER-41036}
\email{calmet@physics.unc.edu}

%\begin{abstract}
Ê% We show that the previously introduced concept of distance on statistical spaces leads to a %straightforward definition of differential entropy on these statistical spaces. These spaces are %characterized by the fact that their points can only be localized within a certain volume and %exhibit thus a feature of fuzziness. This implies that Riemann integrability of relevant integrals %is no longer secured. Some discussion on the specialization of this formalism to quantum %states concludes the paper.Ê
%\end{abstract}

%\keywords{Differential entropy, mutual information, statistical spaces, minimal length, quantum %states, entropic inequalities.}

\maketitle

\section{Abstract}
 We show that the previously introduced concept of distance on statistical spaces leads to a straightforward definition of differential entropy on these statistical spaces. These spaces are characterized by the fact that their points can only be localized within a certain volume and exhibit thus a feature of fuzziness. This implies that Riemann integrability of relevant integrals is no longer secured. Some discussion on the specialization of this formalism to quantum states concludes the paper.Ê
 \\
 \\
 {\bf keywords:} Differential entropy, mutual information, statistical spaces, minimal length, quantum states, entropic inequalities.
\section{1. Introduction}

Differential entropy is the entropy of a continuous random variable. It is related to the shortest description length and thus similar to the entropy of a discrete random variable. A basic introduction can be found in the book of Cover and Thomas \cite{Cover}. In this paper, we are interested in the concept of shortest description length. Indeed, in a recent paper \cite{tenerif}, we have investigated the case of spaces where points are in fact localized within a certain volume, i.e. they are statistical in nature. The motivation was the existence of a minimal length in physical theories. It was possible to introduce a concept of distance using Fisher information metric on such spaces. In this paper we show that the reasoning leading to the definition of a distance is analogous to the usual introduction of differential entropy in information theory. In our case also, care must be taken of the precise meaning of minimal distance or shortest description length.

In this work we only present an outline of our method. It is structured as follows. In section 2 the basic definitions of differential entropy, Kullback-Leibler distance and mutual information are reminded. The following section is devoted to the concept of distance on statistical spaces as proposed in \cite{tenerif}. Section 4 extends the definitions presented in section 2 to the statistical spaces. In particular, we show that the concept of distance introduced in the preceding section leads to a mutual information function analogous to the usual one. This is the main and new contribution of this paper.   Section 5 offers a discussion of some specific features and outlines some open research tracks. It also contains concluding remarks.Ê

\section{2. Basic definitions}

{\bf Definition 1:} The differential entropy $h(X)$ of a continuous random variable $X$ with a density $f(x)$ is defined as
\begin{eqnarray}
h(X)=-\int_S f(x) \log f(x) dx
\end{eqnarray}
where $S$ is the support set of the random variable.

It is well-known that this integral exists if and only if the density function of the random variables is such that the integrals can be defined. This is related to the issue concerning the precise meaning of minimal distance.
The next step consists in establishing the relation of differential entropy to discrete entropy. Then, one proceeds to define the joint and conditional differential entropy including the entropy of a multivariable distribution. The next step is to introduce the relative entropy and the mutual information functions.
\\

{\bf Definition 2:} The Kullback-Leibler  distance or relative entropy is defined as
\begin{eqnarray}
 D(f || g)&=&  
\int   f \log \frac{f}{g}
\end{eqnarray}
where $f$ and $g$ are two density functions.
\\

{\bf Definition 3:} The mutual information $I(X;Y)$ between two random variables with joint density $f(x,y)$ is defined as
\begin{eqnarray}
 I(X;Y)&=&  
\int   f(x,y) \log \frac{f(x,y)}{f(x) f(y)} dx dy.
\end{eqnarray}

The previous definitions lead to an equation for the mutual information given by:
\begin{eqnarray} \label{mut}
 I(X;Y)&=& D(f(x,y) || f(x) f(y)).
\end{eqnarray}
An important point is that this provides a link between the discrete and continuous cases since the properties of the Kullback-Leibler distance and the mutual information are the same in both cases.
\section{3. Distance on statistical spaces}

We now briefly review the connection between the Kullback-Leibler distance and the Fisher Information matrix. In order to do so, we start from a generalized concept of distance based on the concept of entropy.
It is often useful to introduce a concept of distance between elements
of a more abstract set.  For example, one could ask what is the
distance between two distributions between e.g. the Gaussian and
binomial distributions. It is useful to introduce the concept of
entropy as a mean to define distances. In information theory,
Shannon entropy \cite{Shannon} represents the information content of
a message or, from the receiver point of view, the uncertainty about
the message the sender produced prior to its reception. It is defined
as
\begin{eqnarray}
- \sum_i p(i) \log p(i),
\end{eqnarray}
where $p(i)$ is the probability of receiving the message $i$. The unit
used is the bit. The relative entropy can be used to define a
``distance'' between two distributions $p(i)$ and $g(i)$. The
Kullback-Leibler \cite{Kullback} distance or relative entropy is defined as
\begin{eqnarray}
D(g||p)&=& \sum_i g(i) \log \frac{g(i)}{p(i)}
\end{eqnarray}
where $p(i)$ is the real distribution and $g(i)$ is an assumed
distribution. Clearly the Kullback-Leibler relative entropy is not a
distance in the usual sense: it satisfies the positive definiteness
axiom, but not the symmetry or the triangle inequality axioms. It is
nevertheless useful to think of the relative entropy as a distance
between distributions.

The Kullback-Leibler distance is relevant to discrete sets. It can be
generalized to the case of continuous sets. For our purposes, a
probability distribution over some field (or set) $X$ is a
distribution $p:X \in \mathbb{R}$, such that
\begin{itemize}
\item[1.] $\int_X d^4\!x \  p(x)=1$
\item[2.] For any finite subset $S\subset X$, $\int_S d^4\!x \ p(x)>0$.
\end{itemize}
We shall consider families of distributions, and parameterize them by a
set of continuous parameters $\theta^i$ that take values in some open
interval $M \subseteq \mathbb{R}^4$. We use the notation $p_\theta$ to
denote members of the family. For any fixed $\theta$, $p_\theta: x
\mapsto p_\theta(x)$ is a mapping from $X$ to $\mathbb{R}$. We shall
consider the extension of the family of distributions $F=\{ p_\theta|
\theta \in M\}$, to a manifold ${\cal M}$ such that the points $p\in
{\cal M}$ are in one to one correspondence with the distributions
$p\in F$. The parameters $\theta$ of $F$ can thus be used as
coordinates on ${\cal M}$.

The Kullback number is the generalization of the Kullback-Leibler
distance for continuous sets. It is defined as
\begin{eqnarray}
I(g_\theta|| p_\theta)&=&  
\int d^4\!x  g_\theta(x) \log \frac{g_\theta(x)}{p_\theta(x)}.
\end{eqnarray}
Let us now study the case of an infinitesimal difference between
$q_\theta(x)=p_{\theta +\epsilon v}(x)$ and $p_\theta(x)$:
\begin{eqnarray}
I(p_{\theta +\epsilon v}||p_\theta)&=&  \int d^4\!x  p_{\theta +\epsilon v}(x) 
\log \frac{p_{\theta +\epsilon v}(x)}{p_\theta(x)}.
\end{eqnarray}
Expanding in $\epsilon$ and keeping $\theta$ and $v$ fix one finds
(see e.g. \cite{rodriguez3}):
\begin{eqnarray}
\! \!  \! I(p_{\theta +\epsilon v}||p_\theta)&=&I(p+\epsilon||p)
\vert_{\epsilon=0}+ \epsilon \ I^\prime(\epsilon)\vert_{\epsilon=0} \\ \nonumber && +
\frac{1}{2}\epsilon^2 \
I^{\prime\prime}(\epsilon)\vert_{\epsilon=0}+{\cal O}(\epsilon^3).
\end{eqnarray}
One finds $I(0)=I^\prime(0)=0$ and
\begin{eqnarray}
I^{\prime\prime}(0)&=& v^\mu \left ( \int_X d^4\!x p_\theta(x) \left  
(\frac{1}{p_\theta(x)} \frac{\partial p_\theta(x)}{\partial \theta^\mu} \right ) \right. \\ \nonumber &&
\ \ \ \ \ \ \ \ \ \ \ \ \  \ \ \ \ \ \ 
  \left. \left
(\frac{1}{p_\theta(x)} 
\frac{\partial p_\theta(x)}{\partial \theta^\nu} \right ) \right) v^\nu.
\end{eqnarray}
We can now identify the Fisher information metric \cite{Fisher} on a
manifold of probability distributions as
\begin{eqnarray}
g_{\mu\nu}=\int_X d^4\!x \frac{1}{p_\theta(x)}  
\frac{\partial p_\theta(x)}{\partial \theta^\mu}  
\frac{\partial p_\theta(x)}{\partial \theta^\nu}.
\end{eqnarray}
It has been show that this matrix is a metric on a manifold of
probability distributions, see e.g.  \cite{rodriguez1}.

In \cite{tenerif} we have shown that using the concept of relative entropy, one can introduce a concept of a distance, equivalent to the Kullback-Leibler distance, on statistical spaces.

{\bf Definition 4:}  Distance on statistical spaces between two ``points'' $p_{\theta^\mu}(x^\mu)$ and $q_{{\theta'}^\mu}(x^\mu)$ Ê
\begin{eqnarray}
I(q_{{\theta'}^\mu}(x^\mu)||p_{\theta^\mu}(x^\mu) )  &=& 
\int d^4\!x   q_{{\theta'}^\mu}(x^\mu)\log \frac{q_{{\theta'}^\mu}(x^\mu)}{p_{\theta^\mu}(x^\mu)}.
\nonumber \\  && \! \! \! \! \! \!
\end{eqnarray}
The metric on the manifold of
distributions is given locally by
\begin{eqnarray} \!\!
g_{\mu\nu} &=&  \int_X d^4\!x p_\theta(x) \left 
(\frac{1}{p_\theta(x)} \frac{\partial p_\theta(x)}{\partial 
\theta^\mu} \right )
 \left  
(\frac{1}{p_\theta(x)} 
\frac{\partial p_\theta(x)}{\partial \theta^\nu} \right )
\nonumber \\  && 
\end{eqnarray}
and corresponds to the Fisher information matrix.  The distance
between two points $A^\mu$ and $B^\nu$ on the manifold is given by
$d(A^\mu, B^\nu)=\sqrt{g_{\mu\nu}A^\mu B^\nu}$. It was also shown in \cite{tenerif} that a Lorentzian metric can be generated in certain cases of physical relevance. 

\section{4. Mutual information on statistical spaces}

We follow the presentation of differential entropy proposed in chapter 9 of \cite{Cover}, The definitions (1) to (4) extends directly to the case of statistical spaces.

It follows that the properties of the differential entropy, relative entropy and mutual information can be carried out to our framework. In particular, the following relevant theorems still hold since their proofs are based on definitions  (1) to (4) or on the Jensen's inequality \cite{Cover}.Ê

Theorem 1:  \begin{eqnarray}
D(f||g)\ge 0,
\end{eqnarray}
with equality if and only if $f = g$ almost everywhere.Ê
\\

Theorem 2: \\
Chain rule for differential entropyÊ
\begin{eqnarray}
h(X_1,X_2,..., X_n) = \sum^n_{i=1} h(X_i|X_1,X_2,...,X_{i-1}).
\end{eqnarray}
In this case also a corollary of this theorem leads to:Ê
\\

Theorem 3: \begin{eqnarray}
h(X_1,X_2,...,X_{n}) \le \sum h(X_i) 
\end{eqnarray}
with equality if and only if $X_1, X_2, É.,X_n$ are independent.Ê

This leads trivially to the Hadamard's inequality \cite{Cover}. This inequality together with the previous theorems enables to prove that a number of determinant inequalities can be derived from information theoretic inequalities. They can be found in chapter 16 of \cite{Cover}. Ê

It seems that most of the concepts valid for differential entropy in the usual formalism can be applied straightforwardly to statistical spaces. This is mostly right. There is however a distinction when it comes to the relation of differential entropy to discrete entropy. In the usual formalism, the differential entropy of a discrete random variable can be considered to be  infinity. This agrees with the idea that the volume of the support set of a discrete random variable is zero \cite{Cover}.

We are now in a framework where this assumption is no longer valid since the existence of a minimal length forbids a zero support set.

A consequence is that it is no longer proven whether the entropy of a n-bit quantization of a continuous random variable $X$ is approximately $h(X) + n$. Indeed the Riemann integrability of the density function apparently no longer holds. A full investigation of this question is still required.Ê

\section{5. Differential Entropy and Dynamics of Uncertainty}
This concluding section is devoted to a brief discussion of the link between differential entropy, dynamics and the inequalities asserting the degree of uncertainty of the concept of information. This link is not new. Indeed, the definition of Shannon entropy mixes both uncertainty and information measure. Differential entropy can then be seen as an assessment of the uncertainty on the knowledge of the information contents of a system.

This paper is the third in a series defining first the concept of metric on a statistical space-time \cite{tenerif} and then introducing the concept of dynamics in the Fisher Information Metric \cite{Calmet }. A new contribution in this paper is to show that the definition of the metric on a statistical space-time allows to define the same expression for the mutual information $I(X;Y)$ as given in eq. (\ref{mut}). As long as it is not required to specify the density distributions there is a straightforward transcription from the usual macroscopic formalism. This is why the main definitions taken from \cite{Cover} are valid in both formalisms.

These definitions enable in the classical case to prove a series of inequalities that are mostly based upon the Riemann integrability of the integral of $f(x) \log x$ where $f(x)$ is the probability density function. Riemann integrability implies a well-defined concept of limit. In statistical spaces we no longer have the simplifying assumption that the limit of the support set of the random variable is zero. As already mentioned this Riemann integrability is among the problems to be investigated.Ê

An even more interesting question is the case of quantum states. We have borrowed the title of this section from a recent paper \cite{Garba} where the dynamics of uncertainty for quantum states is thoroughly investigated.

It is well-known that the quantum equivalent of a bit is a qubit whose states are vectors in a two-dimensional Hilbert space. It is also well-known that quantum systems have vanishing von Neumann entropy. This means that a complete information on the state of a system is presumed. On the other side, the Shannon entropy is interpreted as a probability distribution. The adequacy of either the von Neumann or Shannon entropy for quantum states has been the topic of several studies, see for instance \cite{Mana}. We obviously fully agree with the option adopted in \cite{Garba} to rely on differential entropy to investigate the density evolution.

In this paper Garaczewski pays special attention to various entropic inequalities including those briefly mentioned above. In an application example, he extends the basic features of his formalism to a so-called information dynamics due to the Schr\"odinger model of evolution of wave packets. He also points out that Shannon differential entropy has been used for years in the formulation of entropic versions of Heisenberg-type indeterminacy relations. Our reference to \cite{Garba} and the reasons to give a brief overview of its goals are motivated by the fact that although we do not specify our formalisms to quantum states, this is an obvious goal of our approach. Indeed, a big advantage is that we can use Gaussian distributions as density functions which are realistic models for quantum states as used in quantum computing. Moreover, we have shown in \cite{Calmet } how to easily introduce dynamics into the formalism. This is achieved by methods known to every particle physicist and consists in imposing symmetries. We are thus apparently well-equipped to investigate entropic inequalities for quantum states. This is part of our agenda of problems to investigate. It must be noted that until this investigation is completed, caution must be taken not to presume any result within our formalism. Indeed, we have no simplifying hypothesis such as the vanishing of the support set of the density functions and thus can only guess that we will be able to prove inequalities.
\\

\end{document}